\documentclass[12pt]{article}
\usepackage{graphicx}
\begin{document}

\title{Counterfactuals in Quantum Mechanics}

 \maketitle \vskip 20 pt

Counterfactuals in quantum mechanics appear in discussions of a)
nonlocality, b) pre- and post-selected systems, and c)
interaction-free measurements. Only the first two issues are related
to counterfactuals as they are considered in the general
philosophical literature:
\begin{quotation}
{\em If it were that $\cal A$, then it would be that $\cal B$.}
\end{quotation}
The truth value of a counterfactual is decided by the analysis of
similarities between the actual and possible counterfactual worlds
\cite{Lewis1}.

The difference between a counterfactual (or counterfactual
conditional) and a simple conditional: {\em If  $\cal A$, then $\cal
B$,} is that in the actual world $\cal A$ is not true and we need
some ``miracle'' in the counterfactual world to make it true. In the
analysis of counterfactuals out of the scope of physics, this
miracle is crucial for deciding whether $\cal B$ is true. In
physics, however, miracles are not involved. Typically:
\begin{quotation}
 $\cal A$ : {\em  A measurement  $\cal M$ is
performed}
\end{quotation}
\begin{quotation} $\cal B$: {\em The outcome of $\cal M$  has property $\cal
P$.}
\end{quotation}
Physical theory does not deal with the questions of which
measurement and whether a particular measurement is performed?
Physics yields conditionals: ``If  ${\cal A}_i$, then ${\cal
B}_i$''. The reason why in some cases these conditionals are
considered to be counterfactual is that several conditionals with
incompatible premises
 ${\cal A}_i$ are considered with regard to a single system.

The most celebrated example is the Einstein-Podolsky-Rosen (EPR)
argument in which incompatible measurements of the position or,
instead, the momentum of a particle are considered. Stapp has
applied a formal calculus of counterfactuals to various EPR-type
proofs \cite{Stapp0,Stapp1} and in spite of extensive criticism
\cite{StappSkyrms,StappRedhead,StappClifton,StappMermin,StappUnruh,StappShimStein},
 continues to claim that the nonlocality of quantum mechanics can be
proved without the assumption ``reality'' \cite{Stapp2}.

Let me give here just the main point of this controversy.  Stapp
provides elaborate arguments in which an {\it a priori uncertain}
outcome of a measurement of $O$ in one location might depend on the
measurements performed on an entangled quantum particle in another
location. But if {\it anything} is different in a counterfactual
world, the outcome of the measurement of $O$ need not be the same as
in the actual world. The core of the difficulty is this randomness
of the outcomes of quantum measurements. The formal philosophical
analysis of counterfactuals which uses similarity criteria,
presupposes that in a counterfactual world which is identical to the
actual world in all relevant aspects up until  the measurement of
$O$, the outcome has to be the same. Thus, Stapp's analysis tacitly
adopts the {\em counterfactual definiteness}
\cite{StappSkyrms,StappRedhead} which is essentially equivalent to
``reality'' or hidden variables and which is absent in the
conventional quantum theory.

Important examples of quantum counterfactuals are {\it elements of
reality}. Consider the following {\it definition} \cite{ER}:
 \begin{quotation}
   If we can {\em infer} with certainty that the result of
   measuring at time $t$ of an observable $O$ is $o$, then, at time
   $t$, there exists an element of reality $O=o$.
\end{quotation}
If we consider several elements of reality which cannot be verified
together, we obtain counterfactuals. A celebrated example is the
Greenberger-Horne-Zeilinger (GHZ) entangled state of three
spin-$1\over 2$ particles \cite{GHZ,GHZMermin}:
\begin{equation}
|\Psi\rangle = {1\over \sqrt
  2}(|{\uparrow}\rangle_A|{\uparrow}\rangle_B|{\uparrow}\rangle_C -
|{\downarrow}\rangle_A|{\downarrow}\rangle_B|{\downarrow}\rangle_C)
.
\end{equation}
We consider spin component measurements of these three particles in
the $x$ and $y$ directions. The counterfactuals (the elements of
reality) have a more general form than merely ``the value of $O$ is
$o$'', they are properties of a set of three measurements:
 \begin{eqnarray}
\nonumber
\{{\sigma_A}_x\} \{{\sigma_B}_x\} \{{\sigma_C}_x\} = -1 ,\\
\{{\sigma_A}_x\} \{{\sigma_B}_y\} \{{\sigma_C}_y\} = 1 ,\\
\nonumber \{{\sigma_A}_y\} \{{\sigma_B}_x\} \{{\sigma_C}_y\} = 1 , \\
\nonumber\{{\sigma_A}_y\} \{{\sigma_B}_y\} \{{\sigma_C}_x\} = 1 .
\end{eqnarray}
Here $\{{\sigma_A}_x\}$ signifies the outcome of a measurement  of
$\sigma_x$ of particle $A$ etc. Since one cannot measure  for the
same particle both $\sigma_x$ and $\sigma_y$ at the same time, this
is a set of counterfactuals. It is a very important set because no
local hidden variable theory can ensure such outcomes with
certainty; there is no solution for the set of equations $(2)$.

Lewis's theory of counterfactuals is asymmetric in time
\cite{Lewis2}. The counterfactual worlds have to be identical to the
actual world during the whole time before $\cal A$, but not after.
This creates difficulty in applications of counterfactuals to
physics and especially to quantum mechanics because ``before'' and
``after'' are not absolute concepts. Different Lorentz observers
might see different time ordering of measurements performed at
different places. Finkelstein \cite{Finkel} and Bigaj \cite{Bigaj}
have attempted to  define time asymmetric counterfactuals to
overcome this difficulty. But in my view, the time asymmetry of
quantum counterfactuals is an unnecessary burden \cite{tsqc}. We
{\it can} consider a time symmetric (or time neutral) definition of
quantum counterfactuals.

The general strategy of counterfactual theory is to find
counterfactual worlds closest to the actual world. In the standard
approach, the worlds must be close only before the measurement. In
the time-symmetric approach, the counterfactual worlds should be
close to the actual world both before and after the measurement at
time $t$. Quantum theory allows for a natural and non-trivial
definition of ``close'' worlds as follows: {\it all outcomes of all
measurements performed before and after the measurement of $O$ at
time $t$ are the same in the actual and counterfactual worlds.}

A peculiar example of time symmetric counterfactuals is the {\em
three box paradox} \cite{3box}. Consider a single particle prepared
at time $t_1$ in a superposition of being in three separate boxes:

\begin{equation}
|\Psi_1\rangle = {{1\over \sqrt
  3}}(|A\rangle + |B\rangle + |C\rangle) .
\end{equation}
At a later time $t_2$ the particle is found in another
superposition:

\begin{equation}
|\Psi_2\rangle = {{1\over \sqrt
  3}}(|A\rangle + |B\rangle - |C\rangle).
\end{equation}

For this particle, a set of counterfactual statements, which are
{\em elements of reality} according to the above definition,  is:
\begin{eqnarray}
\nonumber{\rm \bf P}_A = 1 ,\\
{\rm \bf P}_B = 1 .
\end{eqnarray}
Or, in words: if we open box $A$, we find the particle there for
sure; if we open box $B$ (instead), we also find the particle there
for sure.

Beyond these counterfactual statements, there are numerous
manifestations of the claim that in some sense, this single particle
is indeed in two boxes simultaneously. A single photon which
interacts with this particle  scatters as if there are two
particles: one in $A$ and one in $B$, but two or more photons do not
``see'' two particles. Many photons see this single particle as two
particles if the photons interact weakly with the particle. Indeed,
there is a useful theorem which says that if a strong measurement of
an observable $O$ yields a particular outcome with probability 1,
(i.e. there is an element of reality) then a weak measurement yields
the same outcome. Sometime this is called a {\it weak-measurement
element of reality} \cite{WMER}. The outcomes of weak measurements
are {\it weak values}:
\begin{eqnarray}
\nonumber ({\rm \bf P}_A)_w = 1 ,\\
({\rm \bf P}_B)_w = 1 .
\end{eqnarray}
Contrary to the set of counterfactuals above, the weak measurements
can be performed simultaneously both in box $A$ and box $B$. Thus,
the existence of counterfactuals helps us to know the outcome of
real (weak) measurement.

The three-box paradox and other time-symmetric quantum
counterfactuals have raised a significant controversy
\cite{SS,KastSHPMP,KastSHPMP,VaSHPMP,Kast3box,ER,Kastphil,KastrepV,
Morh,Kirk,Ravon}. It seems that the core of the controversy is that
quantum counterfactuals about the results of measurements of
observables, and especially ``elements of reality'' are understood
as attributing values to observables which are not observed. But
this is completely foreign to quantum mechanics. Unperformed
experiments have no results! ``Element of reality'' is just a
shorthand for describing a situation in which we know with certainty
the outcome of a measurement {\it if} it is to be performed, which
in turn helps us to know how weakly coupled particles are influenced
by the system. Having  ``elements of reality'' does not mean having
values for observables. The semantics are misleading since
``elements of reality'' are not ``real'' in the ontological sense.

An attempt to give counterfactuals some ontological sense, at the
cost of placing artificial constraints on the context in which
counterfactuals are considered, was made by  Griffiths \cite{Grif}.
He showed that counterfactuals have no paradoxical features when
only {\it consistent histories} are considered. Another recent step
in this direction are quantum counterfactuals in very restrictive
``measurement-ready'' situations \cite{Miller}.

 Penrose \cite{Pen} used the term
``counterfactuals'' in a very different sense:
\begin{quote}
   {\em Counterfactuals} are things that might
  have happened, although they did not in fact happen.
\end{quote}
In interaction-free measurements \cite{IFM}, an object is found
because it might have absorbed a photon, although actually it did
not. This idea has been applied to  ``counterfactual computation''
\cite{Jos}, a setup in which the outcome of a computation becomes
known in spite of the fact that the computer did not run the
algorithm (in case of one particular outcome \cite{CFCV}).

 In the framework of the Many-Worlds
Interpretation, Penrose's ``counterfactuals'' are counterfactual
only in one world. The physical Universe incorporates all worlds,
and, in particular,  the world in which Penrose's ``counterfactual''
is actual, the world in which the ``counterfactual'' computer
actually performed the computation.

This work has been supported in part by the European Commission
under the Integrated Project Qubit Applications (QAP) funded by the
IST directorate as Contract Number 015848 and by grant 990/06 of the
Israel Science Foundation.

\vskip .5cm Lev Vaidman\hfill\break School of Physics and Astronomy
\hfill\break Raymond and Beverly Sackler Faculty of Exact
Sciences\hfill\break Tel-Aviv University, Tel-Aviv 69978, Israel.

\end{document}